\documentclass[onecolumn,floatfix,aps,nofootinbib,showkeys]{revtex4-2}
\usepackage[dvips]{epsfig}
\usepackage[english]{babel}
\usepackage{bbm}
\usepackage{verbatim}
\usepackage{array}
\usepackage{float}
\usepackage{bm} 
\usepackage{amsmath}
\usepackage{amsfonts}
\usepackage{amssymb}
\usepackage{hyperref}
\usepackage[dvipsnames]{xcolor}
\hypersetup{colorlinks=true,linkbordercolor=Red,linkcolor=Red, citecolor=Red}
\usepackage{indentfirst} %primeiro paragrafo com espaço
\usepackage{float}
\def\p{\mbox{\boldmath$\displaystyle\boldsymbol{p}$}}
\newcommand{\gdualn}[1]{\overset{\:{}^{{}^{\boldsymbol{\neg}}}}{\smash[t]{#1}}}
\def\C{\mbox{\boldmath$\displaystyle\mathbb{C}$}}
\def\R{\mbox{\boldmath$\displaystyle\mathbb{R}$}}

\usepackage{bbold}

\begin{document}

\title{Singular spinors as expansion coefficients of local spin-half fermionic and bosonic fields: On the two-fold Wigner degeneracy}

\author{R. J. Bueno Rogerio$^{1}$} \email{rodolforogerio@gmail.com}
\author{C. H. Coronado Villalobos$^{2}$} \email{ccoronado@utp.edu.pe}

\affiliation{$^{1}$Centro Universitário UNIFAAT \\
Estrada Municipal Juca Sanches, 1050, CEP 12954-070 - Atibaia - SP, Brazil.}
%\affiliation{$^{1}$ Instituto de F\'isica e Qu\'iımica, Universidade Federal de Itajub\'a - IFQ/UNIFEI,
%	Av. BPS 1303, CEP 37500-903, Itajub\'a - MG, Brazil.}

\affiliation{$^{2}$ Universidad Tecnol\'ogica del Per\'u, Lima-Per\'u.}

%\keywords{Singular spinors; Locality; Lounesto classification.}   

\date{\today}

\begin{abstract}
By scrutinizing the singular sector of the Lounesto spinor classification, we investigate the correct definition of the expansion coefficient functions of local fermionic fields within a fully Lorentz covariant theory. As we can observe, a careful definition of the adjoint structure, directed towards local fields, maps singular spinors into class-2 according to a general spinor classification \cite{beyondlounesto}. 

Furthermore, we investigate all the necessary mathematical tools for constructing local fermionic and bosonic fields and provide insights into the physical implications for the other singular classes. Besides, we also show that incorporating \emph{Wigner degeneracy} maintains the rotational symmetry formalism working in general.

\begin{center}
\textit{In memory of Dharam Vir Ahluwalia.}
\end{center}
\end{abstract}

%\pacs{}
%\keywords{}

\maketitle

\section{Introduction}\label{intro}

According to the Lounesto spinor classification, regular spinors are commonly defined as spinors whose scalar and pseudo-scalar bilinear quantities are not always identically equal to zero. These spinors engender the Dirac spinors as its most representative. Nonetheless, regular classes do not exclude all possibilities of spinors, leaving an open door for different types of spinors in which both scalar and pseudo-scalar bilinear quantities vanish identically. These spinors are commonly referred to as singular spinors \cite{lounestolivro}. This class of spinors has not been extensively explored to date and may have significant applications, as can be seen in \cite{out,jrold1,rog1,rog2,elkostates,Ablamowicz:2014rpa,daRocha:2008we, Meert:2018qzk,tipo4epjc,lfrrsingular,lucagold}. Moreover, singular spinors can be further categorized into distinct sub-classes. These sub-classes are distinguished when the axial-vector bilinear quantity is consistently null.

 In such cases, these spinors are referred to as flagpole spinors. Additionally, sub-classes of singular spinors are characterized by vanishing the antisymmetric tensor bilinear quantity. These particular spinors are known as dipole spinors. The flagpole and dipole spinors are significant as they encompass Majorana and Weyl spinors.  

In reference \cite{rodolfonogo}, it is demonstrated that by employing a general regular spinor as the expansion coefficient function of a fermionic quantum field, the imposition of Dirac dynamics leads to a local quantum field. The coefficients are automatically constrained to a subclass referred to as $L_2$ within the class-2 spinors according to Lounesto's classification. Consequently, for other regular spinors belonging to class-2 but not within the $L_2$ subclass and other regular classes, direct attainment of locality and fulfillment of Dirac dynamics are not achieved. This approach directly arises from the implications of Weinberg's no-go theorem and the Lounesto classification. Further discussions can be found in references \cite{dharamreports,cb1,cb2,cb3,cb4}.

In the first part of this paper, we undertake a similar analysis to the one recently developed in \cite{rodolfonogo}, but with a focus on a different type of spinor, singular spinors. Our investigation delves into the general formalism of quantum fields that incorporate singular spinors as expansion coefficient functions, aiming to elucidate all aspects contributing to a local theory. It is noteworthy that none of the singular classes (class-4, class-5, and class-6) accommodate spinors capable of yielding a local theory, and explicit examples can be found in \cite{jcap, tipo4epjc, newfermionsdharam, chengtipo4}. We highlight these new fields also have been extensively explored in the context of mathematical physics \cite{polarform2020,elkopolar,chenggeneral,chenglagrangian}, cosmology \cite{saulo1,saulo2,saulo3}, phenomenology \cite{roldao2023,roldaofermionic,julioperturbative,chengyukawa}, and other related areas. The locality requirement is consistently achieved by imposing parity symmetry at the classical level; however, such imposition does not necessarily compel singular spinors (or quantum fields) to adhere to the Dirac dynamics. Interestingly, the introduction of parity automatically places any singular spinor into class-2 according to \cite{beyondlounesto}, further reinforcing the previous arguments presented in \cite{rodolfonogo} regarding local fields. As we will see, understanding singular spinors as elements describing a degeneracy beyond spin \cite{elkostates}, namely two-fold degeneracy \cite{dharamnpb}, ensures locality and Lorentz invariance. 

In the second part of this paper, we scrutinize a redefinition of the dual structure which has opened doors to new physics \cite{dharamnpb,dharamspinstatistic}, showing the possibility of evading the spin-statistics theorem, so to speak, besides expanding the realm of spin-1/2 particles. Given the redefinition in the dual, it forces the commutative relation among the annihilation and creation operators (bosonic statistics) instead anticommuting relations, usually established for fermions. This fact was recently developed for a specific type of regular spinor \cite{dharamspinstatistic}, bringing new concepts and surprising new results on the subject. What we will do in this work is to extend the procedure to singular spinors and investigate what new physical information is obtained through this new protocol.

What we  report here is a strong consequence of introducing the new adjoint structure, defined in \cite{dharamspinstatistic}, in the singular spinors theoretical framework. Thus, we start by defining a complete set of singular spinors (bearing in mind the Wigner degeneracy), in their most general form, depending on certain phase factors. Next we define the corresponding spin-half quantum fields and then compute the main quantum correlators, in general grounds, without establishing any relation among the creation and annihilation operators. As one can see, the statistics will be fixed by the demand of locality. Once this is done, we are able to calculate the energy of the fields and also the associated propagator.

The paper is organized as follows: in the next section, we introduce the key aspects of singular spinors and the new dual structure. In Section \ref{nonlocalsect}, we delve into the construction of quantum fields.  It becomes evident that singular spinors inherently possess an emergent Lorentz-breaking term encoded in the spin sums. We delve, then, to a description which we believe be necessary for a quantum field constructed with singular spinors. In Sect.\ref{bosonsect} we define the  quantum fields and determine the right relations among creation and annihilation operators. Once such a task is accomplished, we then define the Hamiltonian and the zero point energy.
In the final section we conclude.  

\section{An overview of singular spinors - trial phases}

We define the spinorial components by setting them as eigenstates of the helicity operator, $\vec{\sigma}\cdot\hat{p}$, which can be expressed as $\vec{\sigma}\cdot\hat{p}\; \phi^{\pm}(\boldsymbol{0}) = \pm \phi^{\pm}(\boldsymbol{0})$ and 
consequently the action of $\Theta$ yields $\vec{\sigma}\cdot\hat{p}\; \Theta\phi^{*\;\pm}(\boldsymbol{0}) = \mp\; \Theta\phi^{*\;\pm}(\boldsymbol{0})$, where $\sigma$ stands for the Pauli matrices and the momentum unit vector reads $\hat{p}=(\sin(\theta)\cos(\phi),\sin(\theta)\sin(\phi), \cos(\theta))$ and the operator $\Theta$ stands for the \emph{Wigner time-reversal operator}, which in the spin-$1/2$ representation reads \cite{jcap}
\begin{eqnarray}
\Theta = \left(\begin{array}{cc}
0 & -1 \\ 
1 & 0
\end{array}\right).
\end{eqnarray} 
In the rest frame, the spinorial components read
\begin{eqnarray}\label{components}
\phi^{+}(\boldsymbol{0}) = \sqrt{m}\left(\begin{array}{c}
\cos(\theta/2)e^{-i\phi/2} \\ 
\sin(\theta/2)e^{i\phi/2}
\end{array}\right), \;\;  \phi^{-}(\boldsymbol{0}) = \sqrt{m}\left(\begin{array}{c}
-\sin(\theta/2)e^{-i\phi/2} \\ 
\cos(\theta/2)e^{i\phi/2}
\end{array}\right). 
\end{eqnarray} It can then be introduced a set of singular spinors, $\lambda$, as follows 
\begin{equation}\label{espi}
\lambda^S_{\{+,-\}}(\boldsymbol{0})=\sqrt{m}\left(\begin{array}{c}
\alpha\Theta\phi_L^{-*}(\boldsymbol{0}) \\ 
\beta\phi_L^{-}(\boldsymbol{0})
\end{array} \right),
\; \lambda^S_{\{-,+\}}(\boldsymbol{0})=\sqrt{m}\left(\begin{array}{c}
\alpha\Theta\phi_L^{+*}(\boldsymbol{0}) \\ 
\beta\phi_L^{+}(\boldsymbol{0})
\end{array} \right),
\end{equation} 
and
\begin{equation}\label{nor} 
\lambda^A_{\{+,-\}}(\boldsymbol{0})=\sqrt{m}\left(\begin{array}{c}
-\alpha\Theta\phi_L^{-*}(\boldsymbol{0}) \\ 
\beta\phi_L^{-}(\boldsymbol{0})
\end{array} \right),
\; \lambda^A_{\{-,+\}}(\boldsymbol{0})=\sqrt{m}\left(\begin{array}{c}
-\alpha\Theta\phi_L^{+*}(\boldsymbol{0}) \\ 
\beta\phi_L^{+}(\boldsymbol{0})
\end{array} \right).
\end{equation}
The symbols $\alpha$ and $\beta$ represent trial phase factors in this context. They will be responsible for keeping track of which spinors class the spinor belongs. The lower indices $\lbrace \pm, \mp \rbrace$ denote the helicities of the left and right transforming components, respectively. The $\lambda$ spinors for an arbitrary momentum can be obtained by applying the boost operator, which, since the spinorial components stand for eigenstates of the $\vec{\sigma}\cdot\hat{p}$ operator, may be recast as 
\begin{equation*}
\mathcal{B}_{\pm} = \sqrt{\frac{E+m}{2m}}\left(1\pm \frac{p}{E+m}\right).
\end{equation*}

The constraints for charge-conjugacy\footnote{In which the charge-conjugation operator is defined as $\mathcal{C}=\gamma_2 \mathcal{K}$, with $\mathcal{K}$ representing the complex conjugation operation.} ($\mathcal{C}\lambda = \pm\lambda$) can be summarized as follows: $a)$ $i\beta^{}=\alpha$ for particle spinors, and $b)$ $i\beta^{}=-\alpha$ for anti-particle spinors. This condition automatically places the $\lambda$ spinors in Lounesto class-5. While all eigenspinors of the $\mathcal{C}$ operator belong to class-5, not all spinors within class-5 exhibit charge-conjugacy. Singular spinors do not satisfy the Dirac equation, as previously pbserved in \cite{jcap,glazov}. Nonetheless, even in most general form, the singular spinors fulfill the Klein-Gordon equation, as expected. 

Within the framework of the Dirac dual $\bar{\lambda} = \lambda^{\dag}\gamma_0$, all singular spinors satisfy the condition $\bar{\lambda}\lambda=0$. Back to the specific cases at hand, it is crucial to conduct a thorough examination of the phases $\alpha$ and $\beta$ in order to determine the appropriate class to which the aforementioned singular spinor belongs. The analysis of the Lounesto classification concerning these phases yields significant cases and subcases, as highlighted in \cite{rodolfoconstraints}. Table I presents all the conditions for classifying the singular spinors to provide a comprehensive overview.
\begin{table}[H]
\centering
\begin{tabular}{cccc}
%\multicolumn{4}{c}{\textbf{Singular spinors}}\\
\hline 
\;\;\;\;\;\;\;\;\;\;$\alpha$\;\;\;\;\;\;\;\;\;\; & \;\;\;\;\;\;\;\;\;\;$\beta$\;\;\;\;\;\;\;\;\;\; & \;\;\;\;\;\;\;\;\;\;Class\;\;\;\;\;\;\;\;\;\; & \;\;\;Condition\;\;\; \\ 
\hline 
\hline 
$\Re(\alpha)$ & $\Re(\beta)$ & 4 & $\alpha\neq\beta$ \\ 

$\mathbb{C}$ & $\mathbb{C}$ & 4 & $|\alpha|^2\neq|\beta|^2$ \\

$\mathbb{C}$ & $\Re(\beta)$ & 4 & $|\alpha|^2\neq|\beta|^2$ \\ 

$\mathbb{C}$ & $\Im(\beta)$ & 4 & $|\alpha|^2\neq|\beta|^2$ \\

$\Im(\alpha)$ & $\Im(\beta)$ & 4 & $|\alpha|^2\neq|\beta|^2$ \\ 

$\Im(\alpha)$ & $\Re(\beta)$ & 4 & $|\alpha|^2\neq|\beta|^2$ \\ 

$\Re(\alpha)$ & $\Re(\beta)$ & 5 & $\alpha=\beta$ \\ 

 $\mathbb{C}$ & $\mathbb{C}$ & 5 & $|\alpha|^2=|\beta|^2$ \\
 
$\mathbb{C}$ & $\Re(\beta)$ & 5 & $|\alpha|^2=|\beta|^2$ \\ 
 
$\mathbb{C}$ & $\Im(\beta)$ & 5 & $|\alpha|^2=|\beta|^2$ \\ 

$\Im(\alpha)$ & $\Im(\beta)$ & 5 & $|\alpha|^2=|\beta|^2$  \\ % ou essa relação $\alpha=\beta$, $\alpha=\beta^*$ or 
  
$\Im(\alpha)$ & $\Re(\beta)$ & 5 & $|\alpha|^2=|\beta|^2$ \\ 

0 & $\Re(\beta)$, $\mathbb{C}$ or $\Im(\beta)$ & 6 & - \\ 

$\Re(\alpha)$, $\mathbb{C}$ or $\Im(\alpha)$  & 0 & 6 & - \\ 
\hline 
\hline 
\end{tabular} 
\caption{The phases conditions to classify singular spinors. $\mathbb{C}$ stands for complex, $\Re(\chi)$ denotes a real $\chi$, while $\Im(\chi)$ indicates an imaginary $\chi$.}
\end{table} 

The Dirac adjoint is typically defined by the following structure: $\bar{\psi} = \psi^{\dag}\gamma_0$. Nevertheless, recent discussions on Dirac spinors and locality in \cite{rodolfonogo} have demonstrated that the appropriate definition of the dual structure is achieved through $\bar{\psi}= [\mathcal{P}\psi]^{\dag}\gamma_0$, where the parity operator is defined as $\mathcal{P} = m^{-1}\gamma^{\mu}p_{\mu}$ \cite{speranca}. A careful examination reveals that introducing the parity operator into the dual structure results in all spinors belonging to class-2 within their specific spinor classification \cite{beyondlounesto}. This finding holds as a general result, encompassing all spinors and demonstrating the close connection between the parity operator and class-2 spinors, which is crucial in ensuring locality for quantum fields. While the possibility of defining dual structures with other discrete symmetries is highlighted \cite{rjdual,rrtaka}, their physical meaning is still to be studied.

\subsection{Taking advantage of a freedom in the dual structure}

Looking towards getting a more involved physical scenario, we define the dual structure taking into account the parity operator\footnote{The choice of the parity operator is justified by the physical and mathematical constraints listed in \cite{rrtaka,rjdual}} $\mathcal{P}$ and a new parameter $s$, as recently shown in \cite{dharamspinstatistic}, leading to the following structure for particles $\stackrel{\neg}{\lambda}^{S}_{h}(\textbf{p}) = [\mathcal{P}\lambda^{S}_{h}(\textbf{p})]^{\dag}\gamma_0$, furnishing  
\begin{eqnarray}\label{dualS}
\stackrel{\neg}{\lambda}^{S}_{\{+,-\}}(\textbf{p}) = \sqrt{m}\mathcal{B}_{-}\left(\begin{array}{c}
\beta\Theta\phi_L^{+*}(\boldsymbol{0}) \\ 
-\alpha\phi_L^{+}(\boldsymbol{0})
\end{array} \right)^{\dag}\gamma_0, \quad \stackrel{\neg}{\lambda}^{S}_{\{-,+\}}(\textbf{p}) = \sqrt{m}\mathcal{B}_{+}\left(\begin{array}{c}
-\beta\Theta\phi_L^{-*}(\boldsymbol{0}) \\ 
\alpha\phi_L^{-}(\boldsymbol{0})
\end{array} \right)^{\dag}\gamma_0,
\end{eqnarray} 
and for antiparticles, the dual structure reads $\stackrel{\neg}{\lambda}^{A}_{h}(\textbf{p}) = [-s\mathcal{P}\lambda^{A}_{h}(\textbf{p})]^{\dag}\gamma_0$,  yielding (for the fermiomic case we set $s=+1$) \footnote{At this stage, if one defines the dual structure for the antiparticle spinors under the following replacement $-\mathcal{P}\rightarrow +\mathcal{P}$, the orthonormal relation in \eqref{orto2} will be the same as the one presented in \eqref{orto1}. This freedom, in a suitable physical context, may lead to a relevant change in the statistics. We thank to Prof. Ahluwalia for bringing our attention to this point.}
\begin{eqnarray}\label{dualA}
\stackrel{\neg}{\lambda}^{A}_{\{+,-\}}(\textbf{p}) = \sqrt{m}\mathcal{B}_{-}\left(\begin{array}{c}
-\beta\Theta\phi_L^{+*}(\boldsymbol{0}) \\ 
-\alpha\phi_L^{+}(\boldsymbol{0})
\end{array} \right)^{\dag}\gamma_0, \quad \stackrel{\neg}{\lambda}^{A}_{\{-,+\}}(\textbf{p}) = \sqrt{m}\mathcal{B}_{+}\left(\begin{array}{c}
\beta\Theta\phi_L^{-*}(\boldsymbol{0}) \\ 
\alpha\phi_L^{-}(\boldsymbol{0})
\end{array} \right)^{\dag}\gamma_0.
\end{eqnarray} 
As highlighted in \cite{dharamspinstatistic}, the choice of $s=+1$ corresponds to a fermionic field, while $s=-1$ corresponds to a bosonic field. The introduction of the parameter $s$ in the dual structure has significant implications for the behaviour of annihilation and creation operators, which must satisfy the appropriate anti-commutation or commutation relations.
The introduced dual structure leads to the orthonormality relations\footnote{These spinors do not yet falls whitin class-2, as indicated by $\omega = \pm m(|\alpha|^2-|\beta|^2)$, for $|\alpha|^2\neq |\beta|^2$.}
\begin{eqnarray}
&&\stackrel{\neg}{\lambda}^{S}_{h}(\textbf{p})\lambda^{S}_{h^{\prime}}(\textbf{p}) = +m(|\alpha|^2+|\beta|^2)\delta_{hh^{\prime}},\label{orto1}
\\
&&\stackrel{\neg}{\lambda}^{A}_{h}(\textbf{p})\lambda^{A}_{h^{\prime}}(\textbf{p}) = -m(|\alpha|^2+|\beta|^2)\delta_{hh^{\prime}},\label{orto2}
\\
&&\stackrel{\neg}{\lambda}^{S}_{h}(\textbf{p})\lambda^{A}_{h^{\prime}}(\textbf{p}) = \stackrel{\neg}{\lambda}^{A}_{h}(\textbf{p})\lambda^{S}_{h^{\prime}}(\textbf{p})= 0. \label{orto3}
\end{eqnarray}

%As previously stated, such a dual redefinition forces the spinor to belong to the class-2 of the classification in \cite{beyondlounesto}.
The above structures yield the following spin sums relations
\begin{eqnarray}\label{spinsumS}
\sum_h \lambda^{S}_{h}(\textbf{p})\stackrel{\neg}{\lambda}^{S}_{h}(\textbf{p}) = m \left(\begin{array}{cccc}
|\alpha|^2 & 0 & 0 & -\alpha\beta^* e^{-i\phi} \\ 
0 & |\alpha|^2 & \alpha\beta^* e^{i\phi} & 0 \\ 
0 & \alpha^*\beta e^{-i\phi} & |\beta|^2 & 0 \\ 
-\alpha^*\beta e^{i\phi} & 0 & 0 & |\beta|^2
\end{array} \right),
\end{eqnarray}
and 
\begin{eqnarray}\label{spinsumA}
\sum_h \lambda^{A}_{h}(\textbf{p})\stackrel{\neg}{\lambda}^{A}_{h}(\textbf{p}) = m \left(\begin{array}{cccc}
-|\alpha|^2 & 0 & 0 & -\alpha\beta^* e^{-i\phi} \\ 
0 & -|\alpha|^2 & \alpha\beta^* e^{i\phi} & 0 \\ 
0 & \alpha^*\beta e^{-i\phi} & -|\beta|^2 & 0 \\ 
-\alpha^*\beta e^{i\phi} & 0 & 0 & -|\beta|^2
\end{array} \right),
\end{eqnarray} 
where the subscript $h$ ranges over $\{+,-\}$ and $\{-,+\}$. The spin sums, in a summarized form, can be written as
\begin{eqnarray}
&&\sum_h \lambda^{S}_{h}(\textbf{p})\stackrel{\neg}{\lambda}^{S}_{h}(\textbf{p}) = m[\mathbb{1}_{(\alpha,\beta)} + \mathcal{M}(\alpha, \beta, \phi)],
\\
&&\sum_h \lambda^{A}_{h}(\textbf{p})\stackrel{\neg}{\lambda}^{A}_{h}(\textbf{p}) = -m[\mathbb{1}_{(\alpha,\beta)} - \mathcal{M}(\alpha, \beta, \phi)],
\end{eqnarray}
with a self evident notation for $\mathbb{1}_(\alpha,\beta)$ and $\mathcal{M}(\alpha, \beta, \phi)$. Both spin sums are not Lorentz invariant due to the $\phi$-dependence. Nonetheless, the above results furnish the following \emph{completeness} relation
\begin{eqnarray}
\frac{1}{2m}\sum_{h}[\lambda^{S}_{h}(\textbf{p})\stackrel{\neg}{\lambda}^{S}_{h}(\textbf{p})-\lambda^{A}_{h}(\textbf{p})\stackrel{\neg}{\lambda}^{A}_{h}(\textbf{p})] = \mathbb{1}_{(\alpha, \beta)}.
\end{eqnarray}
We notice that $\mathcal{M}(\pm i, 1, \phi) \rightarrow \mathcal{G}(\phi)$ recovers all the Elko's standard results \cite{mdobook}. Also, if one defines $\alpha =  \pm 1$ and $\beta = 0$, the dipole's spinors  results are automatically recovered \cite{newfermionsdharam}.

Regarding the $\mathcal{M}(\alpha, \beta, \phi)$ operator, we highlight the following interesting properties: $\mathcal{M}(\alpha, \beta, \phi) = -\mathcal{M}(\alpha, \beta, \phi+\pi)$, $\mathcal{M}(\alpha, \beta, \phi)\lambda^{S}_{h}(\textbf{p}) = +\mathbb{1}_{(\beta, \alpha)}\lambda^{S}_{h}(\textbf{p})$ and $\mathcal{M}(\alpha, \beta, \phi)\lambda^{A}_{h}(\textbf{p}) = -\mathbb{1}_{(\beta, \alpha)}\lambda^{A}_{h}(\textbf{p})$.
There is a general result derived from the findings above. Eqs. \eqref{spinsumS} and \eqref{spinsumA} explicitly demonstrate that singular spinors, in their most general form, lack locality. Consequently, solely defining the dual structure by introducing a parity operator does not fully satisfy the requirements for ensuring this necessary physical aspect.    

\section{Non-locality $\times$ Locality}\label{nonlocalsect}

It is possible to define quantum fields based on expansion coefficients performed by singular spinors belonging above 
\begin{eqnarray}\label{campoquantico}
\mathfrak{f}(x) = \int \frac{d^3p}{(2\pi)^3} \frac{1}{\sqrt{2mE(\textbf{p})}}\sum_{h} \bigg[c(\p,h)\lambda^{S}_{h}(\textbf{p})e^{-ip_{\mu}x^{\mu}} + d^{\dagger}(\p,h)\lambda^{A}_{h}(\textbf{p})e^{ip_{\mu}x^{\mu}}\bigg], 
\end{eqnarray}
and the associated dual 
\begin{eqnarray}\label{campoquanticodual}
\stackrel{\neg}{\mathfrak{f}}(x) = \int \frac{d^3p}{(2\pi)^3} \frac{1}{\sqrt{2mE(\textbf{p})}}\sum_{h} \bigg[c^{\dag}(\p,h)\stackrel{\neg}{\lambda}^{S}_{h}(\textbf{p})e^{ip_{\mu}x^{\mu}} + d(\p,h)\stackrel{\neg}{\lambda}^{A}_{h}(\textbf{p})e^{-ip_{\mu}x^{\mu}}\bigg]. 
\end{eqnarray} 
The creation and annihilation operators shall obey the usual fermionic relations
\begin{eqnarray}
\lbrace c(\p,h),c^{\dag}(\p^{\prime},h^{\prime})  \rbrace = (2\pi)^{3} \delta^3(\textbf{p}-\textbf{p}^{\prime})\delta_{hh^{\prime}}, 
\quad  \lbrace c(\p,h),c(\p^{\prime},h^{\prime})\rbrace = 0 =  \lbrace c^{\dag}(\p,h),c^{\dag}(\p^{\prime},h^{\prime})  \rbrace.
\end{eqnarray}
And similar relations shall be assumed for $d(\p,h)$ and $d^{\dag}(\p,h)$ operators.

As it may be verified, singular spinors satisfy only the Klein-Gordon dynamical equation. Therefore, the field is governed by the Klein-Gordon Lagrangian 
\begin{equation}
\mathcal{L}(x) = \partial_{\mu}\stackrel{\neg}{\mathfrak{f}}(x)\partial^{\mu}\mathfrak{f}(x)-m^2 \stackrel{\neg}{\mathfrak{f}}(x)\mathfrak{f}(x).
\end{equation} 
An important observation stemming from equation this observation concerns the mismatch in mass dimensionality between the Standard Model (SM) fermions and singular spinors, which prevents the inclusion of the latter into SM doublets. The equal time quantum correlator between $\mathfrak{f}(x)$ and its momentum conjugated $\mathfrak{p}(x)$ reads 
\begin{eqnarray}\label{correlator2}
\Big\{\mathfrak{f}(\vec{x},t), \mathfrak{p}(\vec{x}\;^{\prime},t)\Big\} =i\delta^3(\vec{x}-\vec{x}\;^{\prime})\mathbb{1}_{(\alpha,\beta)} + i\int \frac{d^3p}{(2\pi)^3}e^{i\vec{p}\cdot(\vec{x}-\vec{x}\;^{\prime})}\mathcal{M}(\alpha, \beta, \phi),
\end{eqnarray} while the Feynman-Dyson is given by 
\begin{align}\label{propagadorfases}
S_{\textrm{FD}}(x^\prime-x)\propto\int\frac{\text{d}^4 p}{(2 \pi)^4}\,
e^{-i p_\mu(x^{\prime\mu}-x^\mu)}
\frac{\mathbb{1}_{(\alpha, \beta)}+\mathcal{M}(\alpha, \beta, \phi)}{p_\mu p^\mu -m^2 + i\epsilon}.
\end{align}
 
Both matrices appearing in the right-hand side of equations \eqref{spinsumS} and \eqref{spinsumA}, $\mathbb{1}_{(\alpha,\beta)}$ and $\mathcal{M}(\alpha,\beta, \phi)$, are invertible and hold the following properties: $\mathbb{1}^2_{(\alpha,\beta)}=diag(|\alpha|^4, |\alpha|^4, |\beta|^4, |\beta|^4)$ and $\mathcal{M}^2(\alpha,\beta,\phi) = diag(|\alpha|^2, |\alpha|^2, |\beta|^2, |\beta|^2)$. Nonetheless, $\det[\mathbb{1}_{(\alpha,\beta)} \pm \mathcal{M}(\alpha, \beta, \phi)] = 0$, suggesting, then, a careful analysis. The above results highlight one of the main features of the singular spinors entering as expansion coefficients of a quantum field: the quantum field is not naturally local. Some examples and mathematical tools concerning singular spinors and propagators can be readily seen in \cite{propagatormpla,propagatorbsm}. In sharp contrast to the case of regular spinors, introducing parity in defining the dual structure does not fully resolve the locality issue. 

To overcome this problem, until recently, we strove to identify subsidiary conditions (essentially a form of freedom) in defining the adjoint structure. One of the mathematical techniques employed for restoring the locality of the field was accomplished through the utilization of particular freedom inherent in the dual structure of spinors, commonly referred to as $\tau$-deformation \cite{rjtau}. This approach allowed us, within well-defined mathematical boundaries, to establish the locality of the quantum field comprehensively for any given singular spinor while simultaneously preserving the entirety of the physical information encapsulated within said spinors. While the mathematical validity of such a mechanism was firmly established, utilizing such a mathematical instrument seemed to need to align seamlessly with the theory, leading to an uneasy approach. 

However, it has recently been revealed that it is possible to eschew the employment of $\tau$-deformation by recognizing the presence of an additional degeneracy inherent in singular spinors, which extends beyond their spin properties \cite{elkostates}. As a result, the \emph{two-fold degeneracy} autonomously emerges and guarantees the locality of the theory \cite{dharamnpb}. We shall depict some key points of this formulation in what follows.

\section{Two-fold Degeneracy}\label{twofoldsection}

The degeneracy beyond spin, mentioned earlier, stems precisely from the very nature of the spinorial representation we are dealing with. Let us elaborate on that. Since we are dealing with peculiar spin $1/2$ representations of the Lorentz group, some care must be taken in defining the quantum field. The spinors presented in \eqref{espi} and \eqref{nor} are composed of dual helicity spinors, in sharp contrast to Dirac spinors, whose helicity of their components is the same (single helicity). This fact may indicate that spin projections along a given axis, say $z$, cannot be taken as a usual label to address this particle's degree of freedom. In other words, for singular spinors, the label $h$ in the quantum field definition shall not be taken as a label for spin projection. This immediately asks how to ensure that the sum in the quantum field definition is being taken over all the possible labels\footnote{Remember that a linear combination in all particle labels is necessary for the quantum field operator to avoid dealing with the highly complex transformation rules of creation and annihilation operators \cite{weinberg1}.}. In this regard, following the reasoning described in Ref. \cite{dharamnpb}, we call attention to the additional self and anti-self conjugated spinors (with respect to the charge conjugation operator): 
\begin{equation}
\rho^S_{\{+,-\}}(\textbf{p})=\sqrt{m}\mathcal{B}_+\left(\begin{array}{c}
\beta^*\phi_R^{+}(\boldsymbol{0}) \\ 
\alpha^*\Theta\phi_R^{+*}(\boldsymbol{0})
\end{array} \right),
\; \rho^S_{\{-,+\}}(\textbf{p})=\sqrt{m}\mathcal{B}_-\left(\begin{array}{c}
\beta^*\phi_R^{-}(\boldsymbol{0}) \\ 
\alpha^*\Theta\phi_R^{-*}(\boldsymbol{0})
\end{array} \right),
\end{equation} 
and
\begin{equation} 
\rho^A_{\{+,-\}}(\textbf{p})=\sqrt{m}\mathcal{B}_+\left(\begin{array}{c}
-\beta^*\phi_R^{+}(\boldsymbol{0}) \\ 
\alpha^*\Theta\phi_R^{+*}(\boldsymbol{0})
\end{array} \right),
\; \rho^A_{\{-,+\}}(\textbf{p})=\sqrt{m}\mathcal{B}_-\left(\begin{array}{c}
-\beta^*\phi_R^{-}(\boldsymbol{0}) \\ 
\alpha^*\Theta\phi_R^{-*}(\boldsymbol{0})
\end{array} \right).
\end{equation}
Following the same steps as developed above, the new dual structures are defined as 
\begin{eqnarray}\label{dualrhoS}
\stackrel{\neg}{\rho}^{S}_{\{+,-\}}(\textbf{p}) = \sqrt{m}\mathcal{B}_{-}\left(\begin{array}{c}
\alpha^*\Theta\phi_R^{+*}(\boldsymbol{0}) \\ 
\beta^*\phi_R^{+}(\boldsymbol{0})
\end{array} \right)^{\dag}\gamma_0, \quad \stackrel{\neg}{\rho}^{S}_{\{-,+\}}(\textbf{p}) = \sqrt{m}\mathcal{B}_{+}\left(\begin{array}{c}
\alpha^*\Theta\phi_R^{-*}(\boldsymbol{0}) \\ 
\beta\phi_R^{-}(\boldsymbol{0})
\end{array} \right)^{\dag}\gamma_0,
\end{eqnarray}
and for the anti-particles
\begin{eqnarray}\label{dualrhoA}
\stackrel{\neg}{\rho}^{A}_{\{+,-\}}(\textbf{p}) = \sqrt{m}\mathcal{B}_{-}\left(\begin{array}{c}
-\alpha^*\Theta\phi_R^{+*}(\boldsymbol{0}) \\ 
\beta^*\phi_R^{+}(\boldsymbol{0})
\end{array} \right)^{\dag}\gamma_0, \quad \stackrel{\neg}{\rho}^{A}_{\{-,+\}}(\textbf{p}) = \sqrt{m}\mathcal{B}_{+}\left(\begin{array}{c}
-\alpha^*\Theta\phi_R^{-*}(\boldsymbol{0}) \\ 
\beta\phi_R^{-}(\boldsymbol{0})
\end{array} \right)^{\dag}\gamma_0,
\end{eqnarray}
The very orthonormal relations as for the $\lambda$ spinors holds for these spinors
\begin{eqnarray}
&&\stackrel{\neg}{\rho}^{S}_{h}(\textbf{p})\rho^{S}_{h^{\prime}}(\textbf{p}) = +m(|\alpha|^2+|\beta|^2)\delta_{hh^{\prime}},\label{ortorho1}
\\
&&\stackrel{\neg}{\rho}^{A}_{h}(\textbf{p})\rho^{A}_{h^{\prime}}(\textbf{p}) = -m(|\alpha|^2+|\beta|^2)\delta_{hh^{\prime}},\label{ortorho2}
\\
&&\stackrel{\neg}{\rho}^{S}_{h}(\textbf{p})\rho^{A}_{h^{\prime}}(\textbf{p}) = \stackrel{\neg}{\rho}^{A}_{h}(\textbf{p})\rho^{S}_{h^{\prime}}(\textbf{p})= 0. \label{ortorho3}
\end{eqnarray}

Computing the spin sums for the $\rho$ spinors, one readily obtains
\begin{eqnarray}\label{spinsumrhoS}
\sum_h \rho^{S}_{h}(\textbf{p})\stackrel{\neg}{\rho}^{S}_{h}(\textbf{p}) = m \left(\begin{array}{cccc}
|\beta|^2 & 0 & 0 & \alpha\beta^* e^{-i\phi} \\ 
0 & |\beta|^2 & -\alpha\beta^* e^{i\phi} & 0 \\ 
0 & -\alpha^*\beta e^{-i\phi} & |\alpha|^2 & 0 \\ 
\alpha^*\beta e^{i\phi} & 0 & 0 & |\alpha|^2
\end{array} \right),
\end{eqnarray}
and
\begin{eqnarray}\label{spinsumrhoA}
\sum_h \rho^{A}_{h}(\textbf{p})\stackrel{\neg}{\rho}^{A}_{h}(\textbf{p}) = m \left(\begin{array}{cccc}
-|\beta|^2 & 0 & 0 & \alpha\beta^* e^{-i\phi} \\ 
0 & -|\beta|^2 & -\alpha\beta^* e^{i\phi} & 0 \\ 
0 & -\alpha^*\beta e^{-i\phi} & -|\alpha|^2 & 0 \\ 
\alpha^*\beta e^{i\phi} & 0 & 0 & -|\alpha|^2
\end{array} \right).
\end{eqnarray} 
Note that these relations are identical to relations \eqref{spinsumS} and \eqref{spinsumA}, up to a sign in the off-diagonal.
Looking towards to summarize the notation, instead of working with 8 spinors, we start defining $\xi(\boldsymbol{0},h)$ spinors, which reads
\begin{eqnarray}
&&\xi(\boldsymbol{0}, 1) = \lambda^S_{\{+,-\}}(\boldsymbol{0}), \quad \xi(\boldsymbol{0}, 2) = \lambda^S_{\{-,+\}}(\boldsymbol{0}),\\
&&\xi(\boldsymbol{0}, 3) = \rho^S_{\{+,-\}}(\boldsymbol{0}), \quad \xi(\boldsymbol{0}, 4) = \rho^S_{\{-,+\}}(\boldsymbol{0}),
\end{eqnarray}
and now we define $\varrho(\boldsymbol{0}, h)$ spinors
\begin{eqnarray}
&&\varrho(\boldsymbol{0}, 1) = \lambda^A_{\{+,-\}}(\boldsymbol{0}), \quad \varrho(\boldsymbol{0},2) = \lambda^A_{\{-,+\}}(\boldsymbol{0}),\\
&&\varrho(\boldsymbol{0}, 3) = \rho^A_{\{+,-\}}(\boldsymbol{0}), \quad \varrho(\boldsymbol{0}, 4) = \rho^A_{\{-,+\}}(\boldsymbol{0}).
\end{eqnarray}
Interestingly enough, such relations above automatically lead to the following results
for the orthonormal relations
\begin{eqnarray}
&&\stackrel{\neg}{\xi}(\p,h)\xi(\p,h^{\prime}) = m(|\alpha|^2+|\beta|^2)\delta_{hh^{\prime}}, \label{ortoF1}
\\
&&\stackrel{\neg}{\varrho}(\p,h)\varrho(\p,h^{\prime}) = -m(|\alpha|^2+|\beta|^2)\delta_{hh^{\prime}}, \label{ortoF2}
\end{eqnarray}
the spin sums reads
\begin{eqnarray}
&& \sum_{h} \xi(\p,h)\stackrel{\neg}{\xi}(\p,h)  = m(|\alpha|^2+|\beta|^2) \mathbbm{1},\label{spinsumFs}
 \\
&& \sum_{h} \varrho(\p,h)\stackrel{\neg}{\varrho}(\p,h)  = -m(|\alpha|^2+|\beta|^2) \mathbbm{1},\label{spinsumFA}
\end{eqnarray}
finally, the completeness relation is given by 
\begin{eqnarray}\label{completeF}
\frac{1}{2m}\sum_{h}\Big[\xi(\p,h)\stackrel{\neg}{\xi}(\p,h)-\varrho(\p,h)\stackrel{\neg}{\varrho}(\p,h)\Big] = (|\alpha|^2+|\beta|^2)\mathbbm{1} .
\end{eqnarray}

The appreciation of $\rho$ and $\lambda$ spinors in a quantum field is the so-called two-fold degeneracy. It brings a degeneracy beyond the spin in describing the particle content degrees of freedom via a quantum field defined with both sets os spinors as expansion coefficients. As a first goal, it becomes evident that there is no need to employ the $\tau$-deformation protocol. 

When considering the two-fold degeneracy, it becomes evident that there is no need to employ the $\tau$-deformation protocol. Bearing in mind the complete set of $\lambda$ and $\rho$ spinors, we define the quantum field as\footnote{We replaced the usual integration element of volume $\frac{d^3p}{(2\pi)^3}\frac{1}{\sqrt{2mE(\p)}}$ by $\frac{d^3p}{(2\pi)^3}\frac{1}{\sqrt{m(|\alpha|^2+|\beta|^2)E(\p)}}$, otherwise the extra multiplicative term, namely $(|\alpha|^2+|\beta|^2)$, would appear in all the expressions.} 
\begin{eqnarray}\label{campoquanticofinal}
\mathfrak{f}(x) =\int\frac{d^3 p}{(2\pi)^3}
\frac{1}{\sqrt{ m(|\alpha^2+|\beta|^2|) E(\p)}}
\bigg[
\sum_{h} {c}(\p,h)\xi(\p,h) e^{-i p\cdot x}+ \sum_{h} d^\dagger(\p,h)\varrho(\p,h) e^{i p\cdot x}\bigg]
\end{eqnarray}
and the associated dual 
\begin{eqnarray}\label{campoquanticodualfinal}
\stackrel{\neg}{\mathfrak{f}}(x) = \int\frac{d^3 p}{(2\pi)^3}
\frac{1}{\sqrt{ m(|\alpha^2+|\beta|^2|) E(\p)}}
\bigg[
\sum_{h} c^\dagger(\p,h)\gdualn{\xi}(\p,h) e^{i p\cdot x}+ \sum_{h} d(\p,h)\gdualn{\varrho}(\p,h)e^{-i p\cdot x}\bigg] 
\end{eqnarray} 
Such redefined quantum fields allow computing the propagator and the equal-time quantum correlator, which now read
\begin{align}\label{propagadorfinal}
S_{\textrm{FD}}(x^\prime-x)=\int\frac{\text{d}^4 p}{(2 \pi)^4}\,
e^{-i p_\mu(x^{\prime\mu}-x^\mu)}
\frac{\mathbb{1}}{p_\mu p^\mu -m^2 + i\epsilon},
\end{align} 
and 
\begin{eqnarray}\label{correlator2final}
\{\mathfrak{f}(t, \vec{x}),\mathfrak{p}_{f}(t, \vec{x}\;^{\prime})\} =i\delta^3(\vec{x}-\vec{x}\;^{\prime})\mathbb{1},
\end{eqnarray}
where
\begin{equation}
\mathfrak{p}_{f}(t,\vec{x}) = \frac{\partial\stackrel{\neg}{\mathfrak{f}}(t,\vec{x})}{\partial t}.
\end{equation}
and the remaining equal-time field-field and momentum-momentum relations are displayed below 
\begin{eqnarray}
\{\mathfrak{f}(t,\vec{x}), \mathfrak{f}(t, \vec{x}\;^{\prime})\}= 0 = \{\mathfrak{p}_{f}(t,\vec{x}),\mathfrak{p}_{f}(t,\vec{x}\;^{\prime})\}, 
\end{eqnarray}
automatically leading to a local theory. 
It is worth noting that the spinors introduced here yield the same positive definite Hamiltonian and zero-point energy as previously defined in \cite{mdobook}.

\section{Local Spin-half bosonic fields}\label{bosonsect}
In this section, we define a new spin-$1/2$ quantum field, built upon the complete set of singular spinors introduced above, playing the role of expansion coefficient functions. Such a procedure is accomplished by investigating a freedom in the spinorial dual structure, leading, then, to a mass-dimension-one field obeying bosonic rather than the fermionic statistic.
As we may see, the new fields provides a local and Lorentz invariant theory with a positively definite Hamiltonian. The study carried out along this section aims to be an extension of what has recently been proposed in \cite{dharamspinstatistic}.

\section{A new physical scenario encoded on the dual structure}\label{secaodual}
As mentioned before, within the framework of the Dirac dual $\bar{\psi} = \psi^{\dag}\gamma_0$, all singular spinors satisfy the condition $\bar{\xi}(\p,h)\xi(\p,h)=0 = \bar{\varrho}(\p,h)\varrho(\p,h)$. Looking towards get a more involved physical scenario, we may explore a freedom on the dual structure, as did in \cite{dharamspinstatistic}. Thus, the aforementioned definition leads to the following structures  
\begin{eqnarray}\label{dualBS}
\stackrel{\neg}{\xi}(\p,h) = [\mathcal{P}\xi(\p,h)]^{\dag}\gamma_0,
\end{eqnarray} 
and the dual structure for $\varrho(\p,h)$ spinors shall be defined as    
\begin{eqnarray}\label{dualBA}
\stackrel{\neg}{\varrho}(\p,h) = s[-\mathcal{P}\varrho(\p,h)]^{\dag}\gamma_0,
\end{eqnarray}  
now we set $s=-1$. 

With a new dual structure at hands, one may compute the orthonormal relations
\begin{eqnarray}
&&\stackrel{\neg}{\xi}(\p,h)\xi(\p,h^{\prime}) = m(|\alpha|^2+|\beta|^2)\delta_{hh^{\prime}}, \label{ortoB1}
\\
&&\stackrel{\neg}{\varrho}(\p,h)\varrho(\p,h^{\prime}) = m(|\alpha|^2+|\beta|^2)\delta_{hh^{\prime}}, \label{ortoB2}
\end{eqnarray}
And the norm vanishes for any other combination. Note that the inclusion of the parameter $s$ brought a crucial modification to the orthonormal relations above, enforcing both to be identical. As we shall see, without it the formalism becomes internally inconsistent, accordingly our purposes. %Locality commutators are not satisfied, positive definite Hamiltonian is not obtained.

In light of the results obtained above, let us examine what other effects the new dual structures can unveil to us. Thus, we proceed to the computation of another significant physical quantity: the spin sums; yielding
\begin{eqnarray}\label{spinsumBS}
\sum_{h} \xi(\p,h)\stackrel{\neg}{\xi}(\p,h)  = m(|\alpha|^2+|\beta|^2) \mathbbm{1},
\end{eqnarray}
and 
\begin{eqnarray}\label{spinsumBA}
\sum_{h}\varrho(\p,h)\stackrel{\neg}{\varrho}(\p,h) = m(|\alpha|^2+|\beta|^2) \mathbbm{1}.
\end{eqnarray} 
Once again, note that, just like the orthonormality relations, the spin sums are also identical. This results combined contrasts with the other findings presented in the literature, where typically the relation in equation \eqref{ortoB2} and \eqref{spinsumBA} carries a global minus sign, when compared with \eqref{ortoF2} and \eqref{spinsumFA}. It is easy to see that spin sums are invariant under Lorentz transformations. 
As we will see later, this crucial change will bring some physical consequences to the (fermionic/bosonic) statistics of the quantum fields, zero-point energy, among others.

The above results furnishes the following \emph{completeness} relation
\begin{eqnarray}
\frac{1}{2m}\sum_{h}\Big[\xi(\p,h)\stackrel{\neg}{\xi}(\p,h)+\varrho(\p,h)\stackrel{\neg}{\varrho}(\p,h)\Big] = (|\alpha|^2+|\beta|^2)\mathbbm{1} .
\end{eqnarray}
We emphasize that the last result above contrasts with the usual fermionic completeness relation, Eq.\eqref{completeF}, because of the ``$+$'' sign rather than a ``$-$'', which is a direct consequence of \eqref{dualBA}. Such a result is in agreement with \cite{dharambosons,dharamspinstatistic}.

\section{The Statistics}\label{estatistica}
Here we introduce a quantum field operator in which the singular spinors displayed above play the role of expansion coefficient functions. The structure of such quantum field operators is essentially the same of the mass-dimension-one quantum fields previously defined in \cite{mdobook}. Such fields reads
\begin{equation}
\mathfrak{b}(x) =
\int\frac{d^3 p}{(2\pi)^3}
\frac{1}{\sqrt{ m(|\alpha^2+|\beta|^2|) E(\p)}}
\bigg[
\sum_{h} {c}(\p,h)\xi(\p,h) e^{-i p\cdot x}+ \sum_{h} d^\dagger(\p,h)\varrho(\p,h) e^{i p\cdot x}\bigg],\label{eq:fieldb}
\end{equation}
and 
\begin{equation}
\gdualn{\mathfrak{b}}(x) =
\int\frac{d^3 p}{(2\pi)^3}
\frac{1}{\sqrt{ m(|\alpha^2+|\beta|^2|) E(\p)}}
\bigg[
\sum_{h} c^\dagger(\p,h)\gdualn{\xi}(\p,h) e^{i p\cdot x}+ \sum_{h} d(\p,h)\gdualn{\varrho}(\p,h)e^{-i p\cdot x}\bigg],
\end{equation}
as its adjoint. Once the expansion coefficients functions only obey the Klein-Gordon equation, and guided by discussion around field's mass dimensionality and dynamics in \cite[See Chapter 9]{mdobook}, the associated lagrangian for the case at hands reads
\begin{equation}
\mathcal{L}(x) = \partial_{\mu}\gdualn{\mathfrak{b}}(x)\partial^{\mu}\mathfrak{b}(x)-m^2\gdualn{\mathfrak{b}}(x)\mathfrak{b}(x).
\end{equation} 

Looking for a way to determine the statistics for the new quantum fields $\mathfrak{b}(x)$ and $\gdualn{\mathfrak{b}}(x)$, we start computing the canonical equal-time (anti-)commutators $[\mathfrak{b}(t,\vec{x}), \mathfrak{p}(t,\vec{x}\;^{\prime})]_{\pm}$, $[\mathfrak{b}(t, \vec{x}), \mathfrak{b}(t, \vec{x}\;^{\prime})]_{\pm}$ and $[\mathfrak{p}_{b}(t,\vec{x}),\mathfrak{p}_{b}(t,\vec{x}\;^{\prime})]_{\pm}$, in which 
\begin{equation}
\mathfrak{p}_{b}(t,\vec{x}) = \frac{\partial\stackrel{\neg}{\mathfrak{b}}(t,\vec{x})}{\partial t}.
\end{equation} 
To keep the development general, we shall not fix the statistics, neither to be fermionic 
\begin{equation}
\left\{c(\p,h),c^\dagger(\p^{\prime},h^{\prime})\right\} =(2\pi)^3 \delta^3(\p-\p^\prime)\delta_{hh^{\prime}}, \quad \left\{c(\p,h),c(\p^{\prime},h^{\prime})\right\} = 0 =
 \left\{c^{\dag}(\p,h),c^\dagger(\p^{\prime},h^{\prime})\right\}, \label{anticomutador}
\end{equation}
and nor bosonic
\begin{equation}
\left[c(\p,h),c^\dagger(\p^{\prime},h^{\prime})\right] =(2 \pi)^3 \delta^3(\p-\p^\prime)\delta_{hh^{\prime}}, \quad \left[c(\p,h),c(\p^{\prime},h^{\prime})\right] = 0 =
 \left[c^{\dag}(\p,h),c^\dagger(\p^{\prime},h^{\prime})\right], \label{comutador}
\end{equation}
and similar relations are also assumed for the remaining operators $d(\p,h)$ and $d^\dagger(\p,h)$. In this context, \eqref{anticomutador} and \eqref{comutador} will be determined based on the requirement of locality.
 
We start computing $[\mathfrak{b}(t,\vec{x}), \mathfrak{p}_{b}(t,\vec{x}\;^{\prime})]_{\pm}$ in a general way
\begin{eqnarray}\label{comu1}
&&[\mathfrak{b}(t,\vec{x}), \mathfrak{p}_{b}(t,\vec{x}\;^{\prime})]_{\pm} = \int \frac{d^3p}{(2\pi)^3}\frac{d^3p^{\prime}}{(2\pi)^3}\frac{1}{\sqrt{m(|\alpha^2+|\beta|^2|)E(\p)}}\frac{1}{\sqrt{m(|\alpha^2+|\beta|^2|)E(\p^{\prime})}}iE(\p^{\prime})\sum_{h}\sum_{h^{\prime}}\times\nonumber\\ 
&&\qquad\qquad\qquad\qquad\quad \Bigg[c(\p,h)c^{\dag}(\p^{\prime},h^{\prime})\xi(\p,h)\gdualn\xi(\p^{\prime},h^{\prime})e^{-ip_{\mu}x^{\mu}+ip^{\prime}_{\mu}x^{\prime\mu}} - d^{\dag}(\p,h)d(\p^{\prime},h^{\prime})\varrho(\p,h)\gdualn\varrho(\p^{\prime},h^{\prime})e^{ip_{\mu}x^{\mu}-ip^{\prime}_{\mu}x^{\prime\mu}} \Bigg]\nonumber\\
&&\qquad\qquad\qquad\qquad \pm\int \frac{d^3p}{(2\pi)^3}\frac{d^3p^{\prime}}{(2\pi)^3}\frac{1}{\sqrt{m(|\alpha^2+|\beta|^2|)E(\p)}}\frac{1}{\sqrt{m(|\alpha^2+|\beta|^2|)E(\p^{\prime})}}iE(\p^{\prime})\sum_{h}\sum_{h^{\prime}}\times\nonumber
\\
&&\qquad\qquad\qquad\qquad\quad \Bigg[c^{\dag}(\p^{\prime},h^{\prime})c(\p,h)\xi(\p,h)\gdualn\xi(\p^{\prime},h^{\prime})e^{-ip_{\mu}x^{\mu}+ip^{\prime}_{\mu}x^{\prime\mu}} - d(\p^{\prime},h^{\prime})d^{\dag}(\p,h)\varrho(\p,h)\gdualn\varrho(\p^{\prime},h^{\prime})e^{ip_{\mu}x^{\mu}-ip^{\prime}_{\mu}x^{\prime\mu}} \Bigg].\nonumber\\
\end{eqnarray}
A careful analysis of the above expression shows that it is non-vanishing only if one assumes the relation \eqref{comutador}, \emph{i.e.}, picking the minus sign. Otherwise, $[\mathfrak{b}(t,\vec{x}), \mathfrak{p}_{b}(t,\vec{x}\;^{\prime})]_{+}=0$ (which is equivalent to $ \{\mathfrak{b}(t,\vec{x}), \mathfrak{p}_{b}(t,\vec{x}\;^{\prime})\}=0$). This result is a direct consequence of the dual structure in \eqref{dualBA}, which forces the field to hold bosonic traces. Thus, bearing in mind the commutative relation and plugging \eqref{spinsumS} and \eqref{spinsumA} into \eqref{comu1}, it provides
\begin{eqnarray}
[\mathfrak{b}(t,\vec{x}), \mathfrak{p}_{b}(t,\vec{x}\;^{\prime})] = i\int\frac{d^3p}{(2\pi)^3}e^{-i\vec{p}(\vec{x}^{\prime}-\vec{x})}\mathbbm{1},
\end{eqnarray}
translating into
\begin{eqnarray}
[\mathfrak{b}(t,\vec{x}), \mathfrak{p}_{b}(t,\vec{x}\;^{\prime})] = i\delta^3 (\vec{x}\;^{\prime}-\vec{x})\mathbbm{1},
\end{eqnarray}
and the remaining equal-time field-field and momentum-momentum relations furnishes 
\begin{eqnarray}
[\mathfrak{b}(t,\vec{x}), \mathfrak{b}(t, \vec{x}\;^{\prime})]= 0 = [\mathfrak{p}_{b}(t,\vec{x}),\mathfrak{p}_{b}(t,\vec{x}\;^{\prime})], 
\end{eqnarray} 
as expected for scalar (bosonic) fields. Note that the fermionic and bosonic fields defined above are essentially the same. They carry exactly the same mathematical structure; however, what imparts distinct physical information to each of them is the relationships brought about by the spin sums, which are a direct consequence of the parameter $s$ of the new dual structures.
 
Now we examine the energy associated with the fields $\mathfrak{b}(x)$ and $\gdualn{\mathfrak{b}}(x)$, in the lights of the results above. The energy density for this case can be displayed as below
\begin{equation}
\mathcal{H} = \mathfrak{p}_{b}(x)\dfrac{\partial\mathfrak{b}(x)}{\partial t}+ \dfrac{\partial\stackrel{\neg}{\mathfrak{b}}(x)}{\partial t}\stackrel{\neg}{\mathfrak{p}}_{b}(x)- \mathcal{L}.
\end{equation}
Thus, the Hamiltonian is given by integrating the energy density over all the 3-space
\begin{eqnarray}\label{hamiltoniana}
H &=& \int d^3x \mathcal{H}, \nonumber\\
&=& \int d^3x\bigg[\partial_0\stackrel{\neg}{\mathfrak{b}}(x)\partial^0\mathfrak{b}(x) - \partial_i\stackrel{\neg}{\mathfrak{b}}(x)\partial^i\mathfrak{b}(x) + m^2\stackrel{\neg}{\mathfrak{b}}(x)\mathfrak{b}(x)\bigg].
\end{eqnarray}
Following the same development as presented in \cite{jcap}, we  find
\begin{equation}
H = \int\frac{\text{d}^3 p}{(2\pi)^3}\frac{1}{m(|\alpha^2+|\beta|^2|)}E(\p)\Bigg[\sum_{h} c^\dagger(\p,h) c(\p,h) \gdualn\xi(\p,h)\xi(\p,h)
+
\sum_{h} d(\p,h) d^\dagger(\p,h)\gdualn\varrho(\p,h)\varrho(\p,h)\Bigg].
\end{equation}
Recalling the orthonormal relations \eqref{ortoB1} and \eqref{ortoB2} above, the last expression above reduces to\footnote{The factor 2 that appears in equation \eqref{h3} is due to the extra degrees of freedom brought by the two-fold degeneracy}
\begin{equation}\label{h3}
H = 2\int\frac{\text{d}^3 p}{(2\pi)^3} E(\p)\Bigg[\sum_{h} c^\dagger(\p,h) c(\p,h) 
+
\sum_{h} d(\p,h) d^\dagger(\p,h)
\Bigg].
\end{equation}
Because the $``+''$ sign, again a consequence of the $s$ in the dual structure, the result above contrasts with results obtained in \cite{jcap,newfermionsdharam}. Next step we make use of the relation \eqref{comutador} looking towards simplify \eqref{h3}, leading to 
\begin{equation}
H= 2H_0 + \sum_{h} \int\frac{\text{d}^3 p}{(2 \pi)^3}
2E(\p) c^\dagger(\p,h) c(\p,h)
+ \sum_{h} \int\frac{\text{d}^3 p}{(2 \pi)^3}
2E(\p) d^\dagger(\p,h) d(\p,h),
\end{equation}
where 
\begin{equation}
H_0 = \delta^3(0)\int\text{d}^3p \; E(\p).
\end{equation}
We highlight here that the zero-point energy differs by a sign from the zero-point energy for the usual fermionic cases \cite{newfermionsdharam,jcap}. Following the same reasoning as presented in \cite{dharamspinstatistic}, for each fermion (or fermionic mass-dimension-one field) there exists a bosonic counterpart of equal mass then the zero-point energies exactly cancel. Up to our knowledge, being a consequence of the formalism.  

Based on the results obtained, we can conclude that the energy is bounded from below when the commutative relations, \eqref{comutador}, are taken into account. Otherwise, the Hamiltonian may not be positive definite. These recent findings contrast with the typical fermionic set-up.

We now move towards the amplitude of propagation computation. We must consider two events, $x$ and $x'$, and then note that the amplitude of propagation from $x$ to $x'$ is given by the following relation
\begin{align}
\mathcal{A}(x\to x^\prime)=\kappa \Big(\underbrace{\langle\hspace{3pt}\vert
\mathfrak{b}(x^\prime)\gdualn{\mathfrak{b}}(x)\vert\hspace{3pt}\rangle \theta(t^\prime-t)
\pm  \langle\hspace{3pt}\vert
\gdualn{\mathfrak{b}}(x) \mathfrak{b}(x^\prime)\vert\hspace{3pt}\rangle \theta(t-t^\prime)}_{\langle\hspace{4pt}\vert \mathfrak{T} (\mathfrak{b}(x^\prime) \gdualn{\mathfrak{b}}(x))\vert\hspace{4pt}\rangle}\Big), \label{amplitude1}
\end{align}
in which the plus sign holds for bosons and the minus sign for fermions. The arbitrary constant $\kappa\in\C$ is to be further determined by a normalisation condition. $\mathfrak{T}$ is the well known time-ordering operator.

As for the two vacuum-expectation-values that appear in $\mathcal{A}(x\to x^\prime) $ we have
 \begin{align}
\langle\hspace{3pt}\vert
\mathfrak{b}(x^\prime)\gdualn{\mathfrak{b}}(x)\vert\hspace{3pt}\rangle  & =\int\frac{d^3p}{(2 \pi)^3}\left(\frac{1}{ m(|\alpha^2+|\beta|^2|) E(\p)}\right)
 e^{-ip\cdot(x^\prime-x)}
 \sum_{h}\xi(\p,h)\gdualn\xi(\p,h), \label{amplitudeP-S}
\\
\langle\hspace{3pt}\vert
\gdualn{\mathfrak{b}}(x) \mathfrak{b}(x^\prime)\vert\hspace{3pt}\rangle 
  & =  \int\frac{d^3p}{(2 \pi)^3}\left(\frac{1}{ m(|\alpha^2+|\beta|^2|) E(\p)}\right)
 e^{ip\cdot(x^\prime-x)}
 \sum_{h}\varrho(\p,h)\gdualn\varrho(\p,h).\label{amplitudeP-A}
\end{align}

The two Heaviside step functions of equation (\ref{amplitude1}) can now be written in their integral representations
\begin{align}
\theta(t^\prime-t) &=  -\frac{1}{2\pi i}\lim_{\epsilon\rightarrow 0^{+}}\int\text{d}\omega
\frac{e^{i \omega (t^\prime-t)}}{\omega- i \epsilon}, \\
\theta(t-t^\prime) &=  -\frac{1}{2\pi i}\lim_{\epsilon\rightarrow 0^{+}}\int\text{d}\omega
\frac{e^{i \omega (t-t^\prime)}}{\omega- i \epsilon},
\end{align}
where $\epsilon, \omega \in\R$. Thus, with such quantities at hands and following similar steps as presented in \cite{chengthesis}, the amplitude of propagation reads 
\begin{eqnarray}\label{propagador666}
\mathcal{A}_{x\rightarrow x^\prime}
= -\kappa \mathop{\mathrm{lim}}\limits_{\epsilon\rightarrow 0^+}
 \int\frac{\mathrm{d}^3p}{(2\pi)^3}\frac{1}{m(|\alpha^2+|\beta|^2|)E(\boldsymbol{p})}
\int\frac{\mathrm{d}\omega}{2\pi i} 
&\bigg[&\frac{\sum_{h}\xi(\p,h)\gdualn\xi(\p,h)}{\omega - i\epsilon}\mathrm{e}^{i(\omega-E(\boldsymbol{p}))(t^\prime -t)}\,\mathrm{e}^{i 
\boldsymbol{p}.(\mathbf{x}^\prime-\mathbf{x})}\nonumber\\
&\pm &
\frac{\sum_{h}\varrho(\p,h)\gdualn\varrho(\p,h)}{\omega - i\epsilon}\mathrm{e}^{-i(\omega-E(\boldsymbol{p}))(t^\prime -t)}\,\mathrm{e}^{i
\boldsymbol{p}.(\mathbf{x}^\prime-\mathbf{x})}\bigg].
\end{eqnarray}

Shifting $\omega \to p_0 = -\omega+E(\p)$ in the first integral and $\omega \to p_0 = \omega- E(\p)$ in the second integral, and taking into account the definition of the spin-sum relations above, 
%%%%%%%%%%%%%%%%%%%%%%%%%%%%%%%%%%%%%%%%%%%%%%%%%%%%%%%%%%%%%%%%%%%%%%%%%%%%%%%
%%%%%%%%%%%%%%%%%%%%%%%%%%%%%%%%%%%%%%%%%%%%%%%%%%%%%%%%%%%%%%%%
\begin{comment}
we can have \eqref{propagador666} shifted into
\begin{eqnarray}\label{general-propagator-form}
&&\mathcal{A}(x\to x^\prime)
=
\nonumber\\ 
&& i\kappa \mathop{\mathrm{lim}}
\limits_{\epsilon\rightarrow 0^+}
\int \frac{\mathrm{d}^4 p}{(2\pi)^4} \frac{1}{2 E(\boldsymbol{p})m} 
\mathrm{e}^{-i p_\mu(x^{\prime\mu} - x^\mu)}
\bigg[ \frac{ \sum_{h}\xi(\p,h)\gdualn\xi(\p,h)}{E(\p) - p_0 - i\epsilon}
\pm
\frac{\sum_{h}[\varrho(-\p,h)\gdualn\varrho(-\p,h)} {E(\p) + p_0 - i\epsilon}\bigg].\nonumber\\
\end{eqnarray}
Defining
\begin{eqnarray}
M(p) = \sum_{h}\xi(\p,h)\gdualn\xi(\p,h),
\\
N(p) = \sum_{h}\varrho(\p,h)\gdualn\varrho(\p,h),
\end{eqnarray}
one may combine the terms in brackets in eq \eqref{general-propagator-form}, furnishing
 \begin{eqnarray}\label{general-propagator-form2}
\mathcal{A}(x\to x^\prime)=  i\kappa \mathop{\mathrm{lim}}
\limits_{\epsilon\rightarrow 0^+}
\int \frac{\mathrm{d}^4 p}{(2\pi)^4} \frac{1}{2 E(\boldsymbol{p})m} 
\mathrm{e}^{-i p_\mu(x^{\prime\mu} - x^\mu)}
\bigg[ \frac{ (M(\p)\pm N(-\p))E(\p)+(M(\p)\mp N(-\p))p_0}{-(p_{\mu}p^{\mu}-m^2+i\epsilon)}\bigg]
\nonumber\\
\end{eqnarray}
Now, with the right spin sums at hands, 
\end{comment}
%%%%%%%%%%%%%%%%%%%%%%%%%%%%%%%%%%%%%%%%%%%%%%%%%%%%%%%%%%%%%%%%%%%%%%%%%%%%%%%
%%%%%%%%%%%%%%%%%%%%%%%%%%%%%%%%%%%%%%%%%%%%%%%%%%%%%%%%%%%%%%%%%%%%%%%%%%%%%%%
by internal consistency we are forced to pick the $``+''$ sign --- which is equivalent to (\ref{comutador}) --- after some mathematical steps, it yields
\begin{align}\label{propagatormdofinalb}
S_{\textrm{FD}}(x^\prime-x)= \int\frac{\text{d}^4 p}{(2 \pi)^4}\,
e^{-i p_\mu(x^{\prime\mu}-x^\mu)}
\frac{\mathbbm{1}}{p_\mu p^\mu -m^2 + i\epsilon}.
\end{align}
the structure of the propagator above is the very same as expected for fermionic mass-dimension-one fields, check for Eq.\eqref{propagadorfinal}, however, now we are handling with spin-half bosons.

\section{Rotation generators for singular spinors}

This section elucidates several fundamental aspects of the proposed construction, focusing on the profound mathematical and physical implications associated with singular spinors and rotational symmetry. Following Weinberg's methodology \cite[page 196]{weinberg1}, we embark on an in-depth investigation into the requisite conditions for rotational symmetry
\begin{equation}
\sum_{\bar s} u_{\bar\ell}(0, \bar{s}) \mathbf{J}_{\bar{s} s}^{(j)}
=\sum_{\ell} {\cal J}_{\bar{\ell}\ell}\, u_{\ell}(0,s), \label{relation1}
\quad \mbox{and}\quad -\sum_{\bar{s}} v_{\bar{\ell}}(0, \bar{s}) \mathbf{J}_{\bar{s} s}^{(j)^{*}}
=  \sum_{\ell} {\cal J}_{\bar{\ell}\ell}\, v_{\ell}(0, s), 
\end{equation}
where $\mathbf{J}^{(j)}$ and ${\cal J}$ are the angular momentum matrices
Bearing in mind the Weyl basis to write $\mathbf{J}$ and ${\cal J}$ matrices in Eq. \eqref{relation1}, where in the $j=1/2$ representation, it reads
\begin{equation}
 \mathbf{J}^{(1/2)} = \frac{1}{2}\boldsymbol{\sigma}, \qquad -\mathbf{J}^{(1/2)^{*}} = \frac{1}{2}\sigma_2\boldsymbol{\sigma}\sigma_2, \label{spinoperator}
\end{equation}
and
\begin{eqnarray}
{\cal J}_{i0} = -\frac{i}{2} \left( \begin{array}{cc}
    \sigma_i &  0\\
    0 & -\sigma_i
\end{array}\right), \qquad {\cal J}_{ij} = \frac{1}{2}\epsilon_{ijk} \left( \begin{array}{cc}
    \sigma_k &  0\\
    0 & \sigma_k
\end{array}\right), \label{spinoperator2}
\end{eqnarray}
in which $\sigma$ stands for the Pauli matrices. 
One must invoke the two-fold degeneracy for singular spinors to compute the correct rotation operators. A straightforward computation evinces the correct $\mathcal{J}$ (in sharp contrast with the usual Dirac's case), and it yields
\begin{eqnarray}
&&\mathcal{J}_{x}= 
\setlength{\extrarowheight}{0.28cm}\frac{1}{2}\left(
\begin{array}{cccc}
 0 & -\frac{|\alpha|^2-|\beta|^2}{ |\alpha|^2+|\beta|^2} & \frac{2\alpha^* \beta^*}{|\alpha|^2+|\beta|^2} & 0 \\
 -\frac{|\alpha|^2-|\beta|^2}{ |\alpha|^2+|\beta|^2} & 0 & 0 & -\frac{2\alpha^* \beta^*}{|\alpha|^2+|\beta|^2} \\
 \frac{2\alpha\beta}{|\alpha|^2+|\beta|^2} & 0 & 0 & -\frac{|\alpha|^2-|\beta|^2}{|\alpha|^2+|\beta|^2} \\
 0 & -\frac{2\alpha\beta}{|\alpha|^2+|\beta|^2} & -\frac{|\alpha|^2-|\beta|^2}{|\alpha|^2+|\beta|^2} & 0 \\
\end{array}
\right),  
\\
&& \mathcal{J}_{y}= \frac{1}{2}\left(
\begin{array}{cccc}
 0 & -i & 0 & 0 \\
 i & 0 & 0 & 0 \\
 0 & 0 & 0 & -i \\
 0 & 0 & i & 0 \\
\end{array}
\right),
\\
&&\mathcal{J}_{z}= 
\setlength{\extrarowheight}{0.28cm}\frac{1}{2}\left(
\begin{array}{cccc}
 -\frac{|\alpha|^2-|\beta|^2}{ |\alpha|^2+|\beta|^2} & 0 & 0 & -\frac{2\alpha^* \beta^*}{|\alpha|^2+|\beta|^2} \\
 0 & \frac{|\alpha|^2-|\beta|^2}{ |\alpha|^2+|\beta|^2} & -\frac{2\alpha^* \beta^*}{|\alpha|^2+|\beta|^2} & 0 \\
 0 & -\frac{2\alpha\beta}{|\alpha|^2+|\beta|^2} & -\frac{|\alpha|^2-|\beta|^2}{|\alpha|^2+|\beta|^2} & 0 \\
 -\frac{2\alpha\beta}{|\alpha|^2+|\beta|^2} & 0 & 0 & \frac{|\alpha|^2-|\beta|^2}{|\alpha|^2+|\beta|^2} \\
\end{array}
\right).
\end{eqnarray}
We call attention to the fact that these indeed satisfy $[\mathcal{J}_x, \mathcal{J}_y] = i\mathcal{J}_z$ and cyclic permutations --- according $\mathfrak{su}(2)$ algebra.
Such rotation operators match with eq.(17) from \cite{dharamnpb} under the following replacement $\alpha =\pm i$ and $\beta=1$. 

An important observation regarding the structure of the $\mathcal{J}$ operators derived above is that the rotation operators must incorporate spinorial phase factors when dealing with highly general (singular) spinor structures. This fact can be directly verified in other structures such as spin sums \eqref{spinsumS}-\eqref{spinsumA} and the propagator \eqref{propagadorfases} (with the corresponding counterparts for regular spinors found in \cite{rodolfonogo}). The utilization of the physical content aligns with a particular symmetry, emphasizing the crucial role of discrete symmetry in determining the fixation of phase factors. For instance, in the case of Dirac spinors \cite{weinberg1}, physical content arises from imposing parity symmetry. In contrast, for Elko spinors, conjugacy under the charge-conjugation operator needs to be enforced \cite{jcap}. Once a physical constraint is satisfied, the phases become fixed, enabling the complete definition of self-consistent physical observables.

\section{Concluding Remarks and outlooks}\label{remarks}

Regarding regular spinors, the locality is achieved by imposing, at the classical level, that the spinors satisfy the Dirac dynamics. However, the framework under investigation reveals that quantum fields constructed using singular spinors as expansion coefficient functions inherently exhibit non-locality. Remarkably, theories originating from a sector where the spinors do not reside in $L_2$ but are defined within a new dual structure can yield a well-behaved final formulation and may describe particle states that deviate from the usual ones, as observed by Wigner. As we know, part of the physical information of a spinor is encoded in its dual structure. If modifications are made to such structures, new physics can emerge.

As one can see, for a complete description of a local theory, the particles and anti-particles described by the mass-dimension-one fields each have four degrees of freedom, that is, a total of eight spinors. In other words, they exhibit a \emph{degeneracy beyond spin}, contrasting with those of Standard Model fermions. As shown, for singular spinors locality is achieved by a dual redefinition and imposing the two-fold degeneracy protocol.

All the results obtained in this work are due to the existing freedom in the spinorial adjoint structure.  What we report here is a direct consequence of that. We introduce a new spin-half quantum field, endowed with mass-dimension-one, constructed from singular spinors written in their most general form, nonetheless, such spinors carry a new dual structure and this redefinition of the dual structure is responsible for bringing certain consequences for statistics, and consequently for locality, energy, and propagator. The new dual structure opens up the possibility for us to access either fermionic or bosonic statistics, depending on the fixation of a certain parameter.
Therefore, the theory built with these quantum fields is local, Lorentz invariant and ensures a positive-definite Hamiltonian if and only if the statistics are bosonic rather then fermionic.

What we developed here, stands for a general result, and it may confirms that the spin-statistics theorem can be evaded by defining new spinorial duals structures, in total agreement with previous results in \cite{dharamspinstatistic}.

These new fields still have an unknown physics and deserve to be explored in detail from a dynamical perspective, in various scenarios such as cosmology and phenomenology.
 
\section{Acknowledgements}
We sincerely thank Professor Dharam Vir Ahluwalia (\emph{in memoriam}) for his invaluable observations and insightful comments during the manuscript writing process. Furthermore, we would like to express our gratitude for his gracious provision of the file ``October17-Wigner-2022'' \emph{Mathematica} notebook, which greatly facilitated the calculation of the rotation operators. 
We would also like to thank Professor Julio Marny Hoff da Silva for the privilege of his revision and for the insightful discussions and observations during the writing stage. We also thank Cheng-Yang Lee for his correspondence and comments on the subject.

\end{document}